\documentclass[aps,pre,twocolumn]{revtex4-1}

\usepackage{bm,amsmath,amssymb,graphicx,stmaryrd,subfigure}

\newcommand{\bN}{{\bf N}}
\newcommand{\bX}{{\bf X}}
\newcommand{\bR}{{\bf R}}

\begin{document}

\title{Programmed buckling by controlled lateral swelling in a thin elastic sheet}

\author{M. A. Dias}
 \email{madias@physics.umass.edu}
\author{J. A. Hanna}
 \email{hanna@physics.umass.edu}
 \author{C. D. Santangelo}
 \email{csantang@physics.umass.edu}
\affiliation{Department of Physics, University of Massachusetts, Amherst, MA 01003}

\date{\today}

\begin{abstract}
Recent experiments have imposed controlled swelling patterns on thin polymer films, which subsequently buckle into three-dimensional shapes. We develop a solution to the design problem suggested by such systems, namely if and how one can generate particular three-dimensional shapes from thin elastic sheets by mere imposition of a two-dimensional pattern of locally isotropic growth. Not every shape is possible.  Several types of obstruction can arise, some of which depend on the sheet thickness. We provide some examples using the axisymmetric form of the problem, which is analytically tractable.  
\end{abstract}

\maketitle

\section{Introduction}

The inhomogeneous growth of thin elastic sheets is emerging as a powerful method for the design of three-dimensional structures from two-dimensional templates \cite{SharonEfrati10,Kim11}.  Most of the associated theory has focused on predicting the buckled shape of a sheet that results from a given imposed pattern of growth or swelling \cite{NechaevVoituriez01,AudolyBoudaoud02,Marder03,BenAmarGoriely05,Efrati09JMPS,Efrati09PRE,Santangelo09}. This work has led to a number of insights, and continues to present challenges for theorists and experimentalists alike. For most applications, however, one knows the desired surface but not the growth pattern that generates it.  In this paper, we pose this reverse buckling problem and solve its axisymmetric form.

The reverse problem has received little attention thus far, and one might mistakenly assume that the solution is trivial. Certainly, it is an elementary exercise in differential geometry to determine the unique metric associated with a surface.  With recourse to through-thickness variations in material properties, extrinsic curvatures may also be programmed to select 
a unique shape. However, real material systems may be more limited.  Thus, we wish to explore the question of which shapes can be made with thin elastic sheets when \textit{only the two-dimensional midsurface metric} can be prescribed. In a practical sense, these are the shapes programmable by encoding a single, spatially-dependent scalar property into a sheet, namely an isotropic swelling ratio\footnote{This follows from the existence of some conformal coordinate system such that the metric is expressible as $\Omega(u,v)\left(du^2 + dv^2\right)$. Such conformal coordinate systems are guaranteed to exist in a neighborhood of any point for sufficiently well-behaved metrics \cite{Chern55}.}.

How could programming the metric of a shape fail to reproduce the shape?  A metric may have families of shapes from which to choose its buckled configuration.  One sees this possibility immediately using a flat metric; a continuum of cylindrical, truncated conical, and other developable immersions exist, but a piece of the plane has lower bending energy than all. Beyond this, the actual metric realized by the sheet often differs from the prescribed metric because accommodating a finite-thickness bending energy may induce an in-plane strain through geometric compatibility conditions.  Finally, there is no guarantee that a given metric can satisfy all of the boundary conditions at the sheet edges.  Such a sheet presumably forms a boundary layer which, while vanishing in the zero thickness limit \cite{Efrati09PRE}, may have nontrivial effects at finite thickness.  

Thus, we wish to know how to prescribe a metric on a sheet of given thickness to produce a desired shape, and what limits there are to the shapes that can be prescribed exactly, up to and including their boundaries.  We will make some progress towards answering these questions in what follows.

\section{Equilibrium Shape Equations}

Consider a flat sheet and multiply all of its metric components by a nonuniform scalar field $\Omega$ that represents the degree of local, isotropic swelling.  The sheet's preferred extrinsic curvature remains zero locally everywhere, but we have prescribed a new ``target'' \cite{MarderPapanicolaou06} metric $\bar{a}_{\alpha\beta}$ (Greek indices range from 1 to 2) on its midsurface.  The actual metric $a_{\alpha\beta}$ adopted by the sheet's midsurface will, in general, not agree with $\bar{a}_{\alpha\beta}$ for sheets of finite thickness.  However, when designing a shape, it is this realized metric that is known.  We suggest that the natural formulation for our design problem is to begin with the realized metric $a_{\alpha\beta}$ of a desired shape in its most convenient coordinate system, and then compute the appropriate target metric $\bar{a}_{\alpha\beta}$.  Finally, we must return to a coordinate system convenient with respect to the flat configuration. What emerges is not only a swelling factor $\Omega$, \textit{if one exists}, but a set of boundaries that may be cut into the unswelled sheet to produce the desired boundaries of the final shape.

In the appendices, we derive our equilibrium equations from a three-dimensional elastic energy, following the procedure of Efrati and co-workers \cite{Efrati09JMPS}. We have adapted their approach to our purpose by using the \textit{realized} metric, $a_{\alpha\beta}$, rather than the target metric $\bar{a}_{\alpha\beta}$, to raise and lower indices as well as to construct a covariant derivative, a measure of integration, and an elastic tensor. As long as the three-dimensional strains are sufficiently small, the difference in our three-dimensional energies is higher than quadratic order in these strains. The result of this change of viewpoint will be a set of equations involving the two curvature invariants of the surface, with an immediate resemblance to well-known equations describing fluid membranes \cite{Seifert97}. 

Consider a thin sheet of thickness $t$ with midsurface immersion $\bX$ and unit normal $\bN$.  The metric and curvature tensors of the surface $\bX$ are the first and second fundamental forms $a_{\alpha\beta} = \partial_\alpha \bX \cdot \partial_\beta \bX$ and $b_{\alpha\beta} = \partial_\alpha\partial_\beta \bX \cdot \bN = -\partial_\alpha\bN \cdot \partial_\beta \bX$.  Two independent invariants of the curvature tensor may be represented using geometric quantities, namely twice the mean and Gaussian curvatures: $2H \equiv b^\alpha_\alpha$ and $2K \equiv b^\alpha_\alpha b^\beta_\beta - b^\alpha_\beta b^\beta_\alpha$.  The prescribed metric $\bar{a}_{\alpha \beta}$ corresponds to the metric of an ``unstretched'' midsurface. The deviation of the realized metric from the prescribed metric defines twice the in-plane strain: $2\varepsilon_{\alpha\beta} \equiv a_{\alpha\beta}-\bar{a}_{\alpha\beta}$.  

In the limit of small thickness, we may write a two-dimensional energy (Appendix \ref{3Dto2Dreduction}):
\begin{eqnarray}\label{2Denergy}
	E &=& \frac{t}{2}\int_{\mathcal{S}} \sqrt{a} \, \mathcal{A}^{\alpha\beta\gamma\delta} \left[ \varepsilon_{\alpha\beta}(\varepsilon_{\gamma\delta} +  \frac{t^2}{3} H b_{\gamma\delta} -\frac{t^2}{4}b_\gamma^\kappa b_{\kappa\delta}) \right.\\
	& & \left. - \frac{t^2}{3} \varepsilon_{\kappa\beta}b_\alpha^\kappa b_{\gamma\delta}  + \frac{t^2}{12}b_{\alpha\beta}b_{\gamma\delta} \right] + \mathcal{O}(t^3 \| b\|^2 \|\varepsilon\|^2, \, t^5 \| b\| ^4) \, ,\nonumber
\end{eqnarray}
where $\| \, \|$ means magnitude and
\begin{equation}\label{2Delastictensor}
	\mathcal{A}^{\alpha\beta\gamma\delta} \equiv \frac{Y}{1+\nu}\left(\frac{\nu}{1-\nu} \, a^{\alpha\beta}a^{\gamma\delta} + a^{\alpha\gamma}a^{\beta\delta}\right)
\end{equation}
is a two-dimensional elastic tensor incorporating an isotropic Young's modulus and Poisson's ratio $Y$ and $\nu$.  The first and last terms in this energy are the stretching and bending terms of Efrati \emph{et al.} \cite{Efrati09JMPS}.  We view $t\| b\|$ as our small parameter, but expect $\|\varepsilon\|$ to be small near equilibrium.  Accordingly, we retain three coupled strain-curvature terms of order $t^3\| b\|^2 \|\varepsilon\| $ dropped in \cite{Efrati09JMPS}, but not terms of order $t^3\| b\|^2 \|\varepsilon\|^2$.  One may ask why such coupled terms should be retained, if when $\varepsilon$ is small they contribute to the energy at higher order.  We will see that the corresponding terms in the Euler-Lagrange equations of equilibrium are independent of $\|\varepsilon\|$, and of the same order as terms arising from variation of the $t^3\| b\|^2$ bending term.  Similarly, the stretching term of order $t\|\varepsilon\|^2$ produces terms of order $t\|\varepsilon\|$ in the equations.

Variation of the immersion of the midsurface (Appendix \ref{2Dvariation}) leads to equations of equilibrium
\begin{eqnarray}
	2B \left[\nabla_\alpha \nabla^\alpha H + 2H\left(H^2-K\right)\right] - s^{\alpha\beta}b_{\alpha\beta}&=&0 \, , \label{normalequil} \\
	\nabla_\alpha s^{\alpha\beta} &=& 0 \, , \label{tangentialequil}
\end{eqnarray}
free smooth boundary conditions\footnote{Our normal force boundary condition (\ref{normalbc}) differs from that of Efrati \emph{et al.} \cite{Efrati09JMPS} but agrees with those of other sources \cite{MullerPC, Koiter66,Niordson85,LL86} in the appropriate limits.}
\begin{eqnarray}
	\left. 2n_\alpha \nabla^\alpha H  + (1-\nu)l_\gamma \nabla^\gamma (b^{\alpha\beta}n_\alpha l_\beta) \, \right|_{\partial \mathcal{S}} &=& 0 \, , \label{normalbc} \\
	\left. n_\beta \left[B\left(2H^2-(1-\nu)K\right)a^{\alpha\beta} +s^{\alpha\beta}\right] \,  \right|_{\partial \mathcal{S}} &=& 0 \, , \label{tangentialbc} \\
	\left. n_\alpha n_\beta \left[ 2\nu Ha^{\alpha\beta} + (1-\nu)b^{\alpha\beta} \right] \,  \right|_{\partial \mathcal{S}} &=& 0 \, , \label{torquebc} 
\end{eqnarray}
and a corner jump condition for free piecewise-smooth boundaries
\begin{equation}\label{cornerbc}
	\left. \llbracket n_{\alpha} l_\beta \rrbracket \left[ 2\nu Ha^{\alpha\beta} + (1-\nu)b^{\alpha\beta} \right]  \, \right|_{\partial \partial \mathcal{S}} \,=\, 0 \, .
\end{equation}
Here, $\nabla_\alpha$ is a covariant derivative constructed from the realized surface metric, $n$ and $l$ are surface tangents normal and tangent, respectively, to the boundary, $\llbracket \, \rrbracket$ denotes a jump in the enclosed quantities, and $B \equiv \frac{Yt^3}{12(1-\nu^2)}$ is a bending modulus.  We have neglected terms of orders $t^3\| b\| ^2\|\varepsilon\|$ and $t\|\varepsilon\|^2$, with the implicit assumption that derivatives do not affect order.  Finally, the ``effective'' stress tensor is given by
\begin{eqnarray}
	s^{\alpha\beta} &\equiv& t  \mathcal{A}^{\alpha\beta\gamma\delta} \varepsilon_{\gamma\delta} + \frac{t^3}{12} \left(H \mathcal{A}^{\alpha\beta\gamma\delta}b_{\gamma\delta} - b^\alpha_\kappa \mathcal{A}^{\kappa\beta\gamma\delta}b_{\gamma\delta} \right)\nonumber \\ \label{stress}
	& & - \frac{t^3}{8}\mathcal{A}^{\alpha\beta\gamma\delta}b_\gamma^\kappa b_{\kappa\delta} \, .
\end{eqnarray}
It should be apparent from this expression that an unstretched midsurface, that is, one free of in-plane strain, does not imply an unstressed finite-thickness sheet.  This extrinsic contribution to the stress is the \emph{only} result of our retention of coupled strain-curvature terms in the energy.  The tensors $t  \mathcal{A}^{\alpha\beta\gamma\delta} \varepsilon_{\gamma\delta} $ and $ \frac{t^3}{12}  \mathcal{A}^{\alpha\beta\gamma\delta}b_{\gamma\delta}$ are the stress and moment tensors of Efrati \emph{et al.} \cite{Efrati09JMPS}; after explicitly raising indices, the latter becomes the bracketed term in the torque boundary condition \eqref{torquebc} and corner condition \eqref{cornerbc}.  The final term in the definition of the effective stress is an analogous application of the elastic tensor to the ``third fundamental form'' $c_{\gamma\delta} \equiv b_\gamma^\kappa b_{\kappa\delta}$.  
%

This stress plays the same role as that of the Lagrange multipliers of Guven \& M\"{u}ller \cite{GuvenMuller08}, whose equilibrium equations for paper coincide with ours when $K=0$.  Actually, the definition of ``stress'' is rather malleable.  Our definition's inclusion of the extrinsic terms that manifest in the divergence-free quantity in \eqref{tangentialequil} seems natural, especially as they arise from variation of the energy with respect to the in-plane strain.  Operationally, any scalar $T$ that does not vary, or varies such that $\delta T = T^{\alpha\beta}\delta a_{\alpha\beta}$, will produce only terms that may be tucked into $s^{\alpha\beta}$.  We note also the absorption of gravitational forces into the Lagrange multipliers in \cite{GuvenMuller08}, reminiscent of the definition of ``dynamic pressure'' in problems involving isochoric fluids.  For more ambiguities, see \cite{Guven04} and the discussion of ``null stresses'' in \cite{Guven06}.

If these equations are to be solved for the six terms $a_{\alpha\beta}$ and $b_{\alpha\beta}$, rather than an explicit immersion $\bX$, they must be supplemented by the Peterson-Mainardi-Codazzi and Gauss equations:
\begin{eqnarray}
	& &\nabla_\alpha b_{\beta\gamma} = \nabla_\beta b_{\alpha\gamma} \, , \label{codazzi} \\
	& &K = a^{\alpha\beta}\left(\partial_\gamma \Gamma^\gamma_{\alpha\beta} - \partial_\alpha \Gamma^\gamma_{\beta\gamma} + \Gamma^\gamma_{\delta\gamma}\Gamma^\delta_{\alpha\beta} - \Gamma^\gamma_{\delta\alpha}\Gamma^\delta_{\beta\gamma}\right). \, \label{gauss}
\end{eqnarray}
The $\Gamma^\alpha_{\beta\gamma}$ are the usual Christoffel symbols.  These auxiliary equations are automatically satisfied by any immersion $\bX$ or $\bX + \delta \bX$.

Many shapes cannot satisfy the boundary conditions (\ref{normalbc}-\ref{torquebc}).  For example, the normal force and torque boundary conditions \eqref{normalbc} and \eqref{torquebc} are incompatible for minimal surfaces of the helicoid-catenoid family.  Such shapes require either a boundary layer or applied boundary forces and torques.

A consequence of the corner condition \eqref{cornerbc} may be observed by bending two adjacent sides of a piece of paper towards each other.  Curvature must vanish ``across'' the sheet at the corner, so the tip remains flat.  

\section{Making shapes}

When the prescribed metric is given and one is solving ``forwards'' for the shape, one must solve the equilibrium equations (\ref{normalequil}-\ref{tangentialequil}) and geometric integrability conditions (\ref{codazzi}-\ref{gauss}) for the six components $a_{\alpha \beta}$ and $b_{\alpha \beta}$.  In the reverse problem, we choose a shape $\bX$ that satisfies two boundary conditions, \eqref{normalbc} and \eqref{torquebc}; integrability is automatically satisfied. After solving the equilibrium equations (\ref{normalequil}-\ref{tangentialequil}), along with boundary condition \eqref{tangentialbc}, for the components of the stress tensor $s^{\alpha \beta}$, we recover \emph{via} \eqref{stress} the target metric $\bar{a}_{\alpha \beta}$ in \textit{whatever buckled coordinate system} we chose for our initial convenience on $\bX$. Finally, we must determine a coordinate transformation back into an appropriate laboratory frame for assigning the swelling factor $\Omega$ to the unbuckled sheet \cite{Dodgson}. 

Though many shapes do not satisfy all of the boundary conditions, in principle only a boundary layer is needed to balance the normal force \eqref{normalbc} and torque \eqref{torquebc} conditions.  Note that this layer may be incorporated into the prescribed swelling factor, so need not share the characteristic width of spontaneously formed layers \cite{Efrati09PRE}. The tangential force condition \eqref{tangentialbc} is more involved. In general, the integration constants of the first-order equation \eqref{tangentialequil} may be insufficient to balance these in-plane forces, which may require global changes in the metric.  Below, we will explore how these boundary conditions affect the construction of axisymmetric shapes, for which we solve the equations of equilibrium analytically.

\subsection{The axisymmetric solution}

Consider an axisymmetric shape parametrized by arc length along a meridional geodesic, so that:
\begin{equation}
	\bX(u,v)=\left(\rho(u)\cos{v}, \, \rho(u)\sin{v}, \, \int_0^u \!\! dy \,\sqrt{1-\left[\partial_{y}\rho(y)\right]^2}\right) \, .
\end{equation}
Thus, $a_{\alpha\beta}dx^\alpha dx^\beta = du^2+\rho^2dv^2$ and $b_{\alpha\beta}dx^\alpha dx^\beta = -\left[\partial_u^2 \rho/\sqrt{1-(\partial_u\rho)^2}\right]du^2+\rho\sqrt{1-(\partial_u\rho)^2}\,dv^2$, with mean curvature $H= \left[1- \partial_u(\rho \partial_u \rho)\right]/\left[2\rho\sqrt{1-(\partial_u\rho)^2}\right]$ and Gaussian curvature $K = -\partial_u^2 \rho/\rho$.

We are free to specify the function $\rho$, as long as $(\partial_u \rho)^2 < 1$.  If we assume a diagonalized, axisymmetric target metric, the equilibrium equations reduce to one differential equation
\begin{equation}\label{eqstress}
\partial_u s^{uu} + s^{uu} \partial_u \ln \left[\rho \sqrt{1-(\partial_u \rho)^2} \right] + g(u) =0 \, ,
\end{equation}
and one algebraic equation
\begin{equation}
s^{vv} = \frac{\partial_u^2 \rho}{\rho\left[1-\left(\partial_u \rho \right)^2\right]} s^{uu} - \frac{g(u)}{\rho\partial_u \rho} \, ,
\end{equation}
where we have defined
\begin{equation}
g(u) \equiv -2B\frac{\partial_u \rho}{\sqrt{1-(\partial_u \rho)^2}} \left(\partial_u^2 H + \frac{\partial_u \rho}{\rho}\partial_u H +2H(H^2-K)\right) \, .
\end{equation}

We can integrate equation (\ref{eqstress}) to yield
\begin{equation}\label{axisymstress}
s^{uu} = \frac{1}{\rho \sqrt{1-\left(\partial_u \rho \right)^2}} \left[C-\int^{u}_{u_b} dy \, g(y) \rho(y) \sqrt{1-\left[\partial_y \rho(y)\right]^2} \right] \, ,
\end{equation}
where $C$ is an integration constant, and $u_b > u$ lies on one boundary of the sheet.  Given the stress tensor, Eq.\ \eqref{stress} is now an algebraic equation for the strain tensor and, thus, the prescribed metric $\bar{a}_{\alpha \beta}$.  

Our coordinates $(u,v)$ are natural for the buckled object, but not for the laboratory.  We must perform a change of variables to a coordinate system convenient for programming an isotropic swelling factor $\Omega(r)$.  A natural choice for axisymmetric shapes is to use cylindrical polar coordinates $(r,\theta)$ and identify $v$ with $\theta$, so that the metric becomes $\Omega(r) (dr^2+r^2 d\theta^2 )$.  The coordinate transformation $u(r)$ is determined by the solution to the differential equation
\begin{equation}
\left[\partial_r u(r)\right]^2 = \frac{\bar{a}_{vv}[u(r)]}{r^2 \bar{a}_{uu}[u(r)]} \, ,
\end{equation}
and the swelling factor by
\begin{equation}
	\Omega(r) = \frac{\bar{a}_{vv}[u(r)]}{r^2}\, .
\end{equation}

\subsection{The axisymmetric boundary conditions}

Our construction relies on our ability to find an $\bX$ that satisfies the free boundary conditions.  Two of these, \eqref{normalbc} and \eqref{torquebc}, are simply two conditions on $\rho(u)$ on the boundaries.  On the boundary $u_b$, the first of these is
\begin{equation}
\rho''(u_b) = \nu \frac{1-\rho'^2(u_b)}{\rho(u_b)} \, ,
\end{equation}
which implies $K(u_b) = - \rho''(u_b)/\rho(u_b) < 0$. The second is
\begin{equation}
\rho'''(u_b) = \frac{(1+\nu+\nu^2)}{\nu} \rho'(u_b) K(u_b) \, .
\end{equation}
These conditions can be easily satisfied using an arbitrarily narrow region near the boundary $u_b$.  If there is another boundary $u_a < u_b$, the same considerations apply there.  

\begin{figure}[h]
\subfigure[~Swelling factor]{
\includegraphics[width=3in]{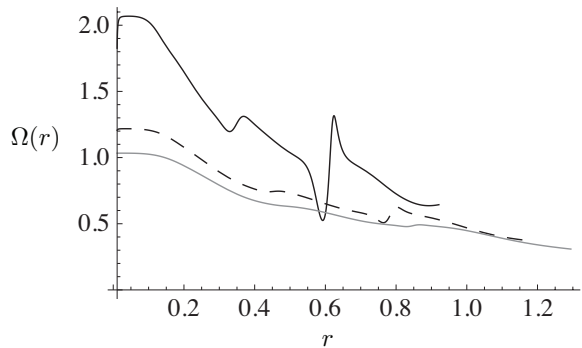}
}
\subfigure[~Ziggurat]{
\includegraphics[width=3in]{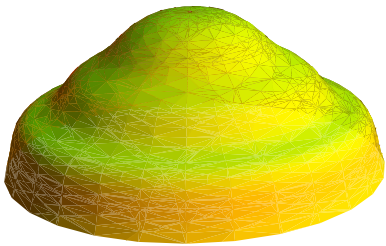}
}
\caption{\label{fig:disk} (Color online) (a) Swelling factor $\Omega(r)$ that swells a disk into the shape in (b) for thicknesses $t=1/100$ (solid black), $t=1/200$ (dashed black), and $t=1/500$ (solid grey). The outer radii needed are $\approx 0.9$, $\approx 1.2$ and $\approx 1.3$, respectively. Lengths are in units of the radial arc length of the final shape.}
\end{figure}
We can also satisfy the boundary condition \eqref{tangentialbc} at $u_b$ by choosing the integration constant
\begin{equation}\label{eqC}
C = - B\left[2H(u_b)^2-(1-\nu)K(u_b)\right]\rho(u_b) \sqrt{1-\rho'(u_b)^2} \, .
\end{equation}
If the surface has only one boundary, this procedure is sufficient.  However, with two boundaries we must also satisfy an integral constraint,
\begin{eqnarray}\label{integraleq}
& &\int_{u_a}^{u_b} du ~g(u) \rho (u) \sqrt{1-\rho '(u)^2} \, =\\
& & - C - B\left[2H(u_a)^2-(1-\nu)K(u_a)\right]\rho(u_a) \sqrt{1-\rho'(u_a)^2} \, .\nonumber
\end{eqnarray}
This is a nonlocal constraint as it involves both boundaries, located at $u_a$ and $u_b$. This global balance may require a global change in $\rho(u)$ to accommodate.  Our procedure, by no means unique, for this accommodation is shown below in the example of the asymmetric annular sheet.

\subsection{Examples}

We start our survey of examples with a topological disk. The swelled shape is shown in Fig.\ \ref{fig:disk}(b) and the swelling factor used to produce it is shown in Fig.\ \ref{fig:disk}(a) for three different sheet thicknesses. This figure was produced using a metric of the form
\begin{eqnarray}\label{eq:crazydisk}
\rho(u) &=& u + A_1 u^4 e^{-10} \left[e^{10 u} - 1\right]\\
& &+A_2 u^5 e^{-10} \left[e^{10 u} - 1\right] - \frac{0.3 u^5}{1+10 u^4}\nonumber\\
& &-\frac{0.3 u^5}{0.01+u^4/[\sin (7 \pi u)/(7 \pi) + 0.5]},\nonumber
\end{eqnarray}
where $A_1$ and $A_2$ were chosen to satisfy the normal force and torque boundary conditions at $u=u_b=1$. We find $A_1\approx -0.829$ and $A_2 \approx -0.723$. Despite its seeming absurdity, Eq.\ (\ref{eq:crazydisk}) underscores the flexibility we have in choosing a metric. Moreover, there is some method to our choice. Since stresses cannot diverge, we require that $\rho(u)$ asymptotically flatten at the center.  It is a straightforward calculation, by expanding the stresses in a power series in $u$, to show that this requires $\rho(0)=0$, $\rho'(0)=1$, $\rho''(0)=0$, $\rho'''(0) < 0$, and $\rho''''(0)=0$. Following the procedure described in the previous two sections, the swelling factor is easily obtained.

The process of choosing coefficients for Eq.\ (\ref{eq:crazydisk}) reveals some of the potential pitfalls of designing a shape. Some $A_1$ and $A_2$ that satisfy the boundary conditions require $|\rho'(u)| > 1$ at one or more places within the sheet.  Moreover, one could find that the resulting prescribed metric is not positive definite everywhere within sheets that are too thick.

\begin{figure}[h]
\subfigure[~Swelling factor]{
\includegraphics[width=3in]{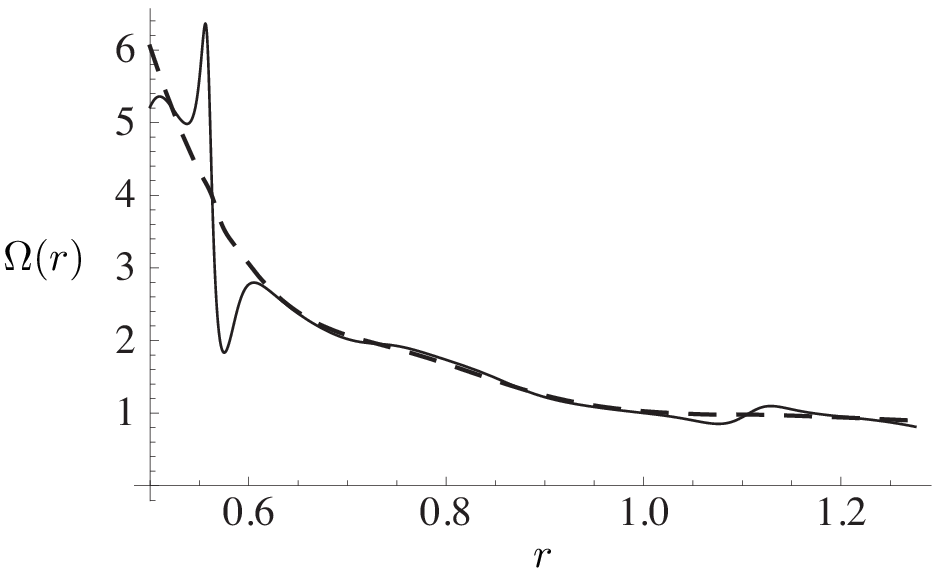}
}
\subfigure[~Sheave]{
\includegraphics[width=3in]{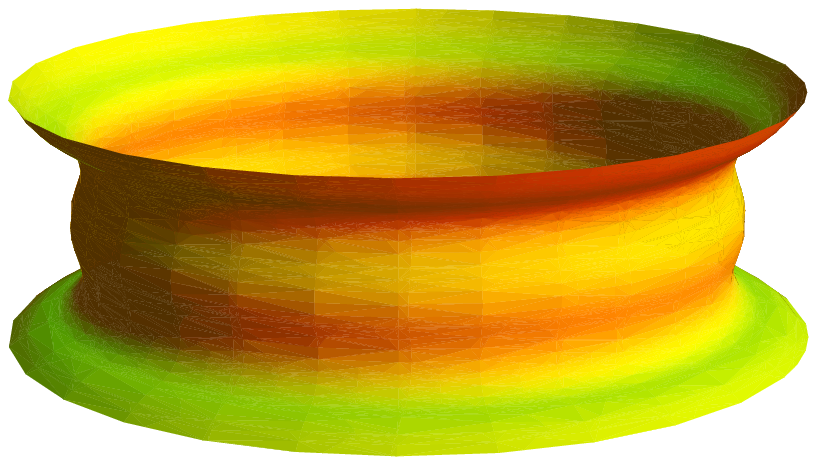}
}
\caption{\label{fig:tube} (Color online) (a) Swelling factor $\Omega(r)$ that swells an annulus with inner radius $r=0.5$ and outer radius $r\approx1.28$ into the shape in (b) for thicknesses $t=1/100$ (solid black) and $t=1/500$ (dashed black). Lengths are in units of the radial arc length of the final shape.}
\end{figure}

We now consider a pair of topological annuli, shapes with two free boundaries.  Satisfying conditions on both boundaries is simple for a surface symmetric about a fixed $u$.  Choosing $C$ to satisfy one boundary, we automatically satisfy the other.  An example is shown in Fig. \ref{fig:tube}.  The metric we use is
\begin{eqnarray}
\rho(u) &=& A_1e^{-10} \left[e^{10 u} - 1\right]+A_2 (1-u)e^{-10} \left[e^{10 u}-1\right]\nonumber\\  \label{eq:crazytube}
& &+ B_1 \left[e^{-10 u} - e^{-10}\right]+B_2 u \left[e^{-10 u}-e^{-10}\right] \\
& & + 1 + \frac{0.4}{5 \pi} \sin (5 \pi u).\nonumber
\end{eqnarray}
We find $A_1 = B_1 \approx 0.205$ and $A_2 = B_2 \approx 1.013$ in order to satisfy the normal force and torque boundary conditions at $u=u_a=0$ and $u=u_b=1$.

\begin{figure}[h]
\subfigure[~Swelling factor]{
\includegraphics[width=3in]{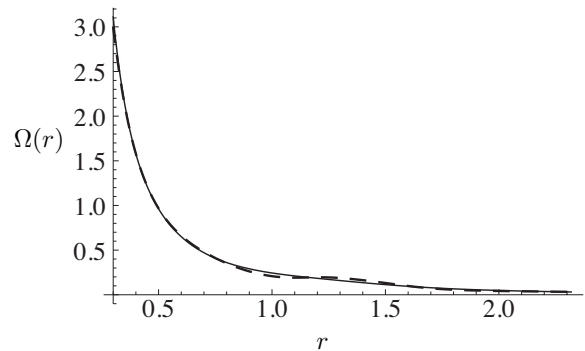}
}
\subfigure[~Compression fitting]{
\includegraphics[width=3in,trim=0in .3in 0in 0in,clip=true]{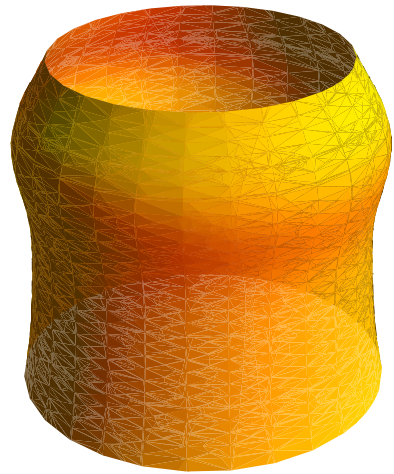}
}
\caption{\label{fig:tubea} (Color online) (a) Swelling factor $\Omega(r)$ that swells an annulus with inner radius $r=0.3$ and outer radius $r\approx 1.81$ into the shape in (b) for thicknesses $t=1/20$ (dashed black) and $t=1/100$ (solid black). Lengths are in units of the radial arc length of the final shape.}
\end{figure}

An asymmetric annulus is significantly more complicated.  It is no longer sufficient to simply choose a metric appropriately on the boundaries, because the integral constraint \eqref{integraleq} depends on the value of $\rho(u)$ throughout the sheet.  After satisfying one boundary, it will generally be impossible to satisfy the other without modifying the metric.  For the example in Fig.\ \ref{fig:tubea}, we use
\begin{eqnarray}
\rho(u) &=& A_1e^{-10} \left[e^{10 u} - 1\right]+A_2 (1-u) e^{-10}\left[e^{10 u}-1\right]\nonumber\\  \label{eq:crazytubeasymmetric}
& &+ B_1 \left[e^{-10 u} - e^{-10}\right]+B_2 u \left[e^{-10 u}-e^{-10}\right] \\
& & - \eta u + 0.5 +\frac{1}{16} e^{-32 (u-3/4)^2},\nonumber
\end{eqnarray}
adjusting $A_1$, $A_2$, $B_1$ and $B_2$ according to the normal force and torque boundary conditions at $u=u_a=0$ and $u=u_b=1$, choosing $C$ to satisfy the tangential force boundary condition at $u=u_b=1$ and the parameter $\eta$ to satisfy the tangential force boundary condition at $u=u_a=0$. We find $A_1\approx-0.044$, $A_2\approx0.029$, $B_1\approx-0.18$ and $B_2\approx0.10$.  Although there is some weak thickness dependence in $\eta$, we find $\eta \approx -0.051$ for thicknesses from $t=1/20$ to $t=1/500$.  There is also a weaker dependence on thickness for $\Omega(r)$ in this example, presumably because this swelled shape has less curvature than the others.

\begin{figure}[h]
\subfigure[~Swelling factor]{
\includegraphics[width=3in]{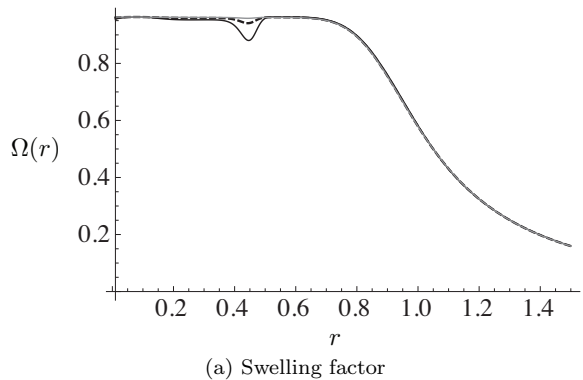}
}
\subfigure[~Drum]{
\includegraphics[width=3in,trim=0in .5in 0in 0in,clip=true]{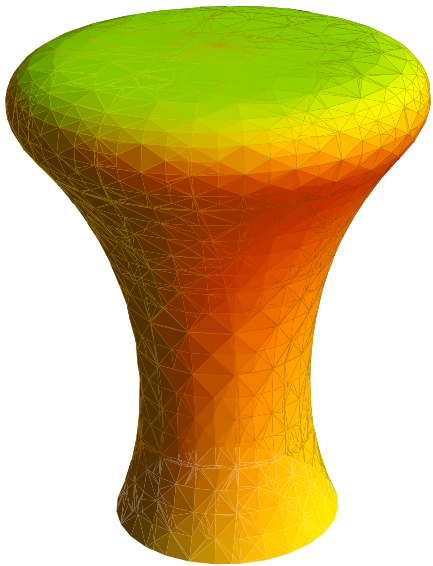}
}
\caption{\label{fig:drum} (Color online) (a) Swelling factor $\Omega(r)$ that swells a disk with outer radius $r\approx 60$ into a ``drum'' with a total radial arc length of $2.75$ units, for thicknesses $t=1/100$ (solid black), $t=1/200$ (dashed black), and $t=1/500$ (solid grey).  We show $\Omega(r)$ only up to $r=1.5$, beyond which it simply approaches zero.}
\end{figure}

Our final example is the disk shown in Fig.\ \ref{fig:drum}, with
\begin{eqnarray}
	\rho(u) & \approx & \left(6.22\times 10^{-11}\right)e^{-10} \left(e^{10 u}-1\right) u^4\nonumber\\
	& &+ \left(-2.1\times10^{-11}\right)e^{-10}\left(e^{10u}-1\right) u^5\nonumber\\
        &   & +\frac{1}{8} \left[(5.966-7.02 u) \text{erf}(7.07 u-6.01) \right.\nonumber\\
        & &+\left(u [1.2 u-2.16]+0.9855\right) \text{erf}(6.67u-6)\\ 
        &   & -\frac{0.13 u^3}{125 u^3+1}+\left(3.39\times10^{-12}\right)u^4+1.2 u^2 \nonumber\\
        & &-1.18 u +6.95 +e^{-44.44 (u-0.9)^2} (0.1 u-0.09)\nonumber\\ 
        &   & \left.-0.56 e^{-50(u-0.85)^2}\right]\nonumber
\end{eqnarray}
This ``drum'' requires a disk of radius $60$ to produce, though it has a total center-to-edge arc length of only $2.75$. This requires significant shrinking, up to a local swelling factor of $\approx 5 \times 10^{-5}$.  This, of course, is another obstruction to swelling a shape: the required swelling factor may be beyond the capabilities of any existing experimental system.  So one can swell the shape of a drum in principle, though perhaps not currently in practice.  We note, however, that conformal transformations of the prescribed metric into some atypical coordinate system may be a way to improve the range of required swelling factors.

\begin{figure}[h]
\includegraphics[width=3in]{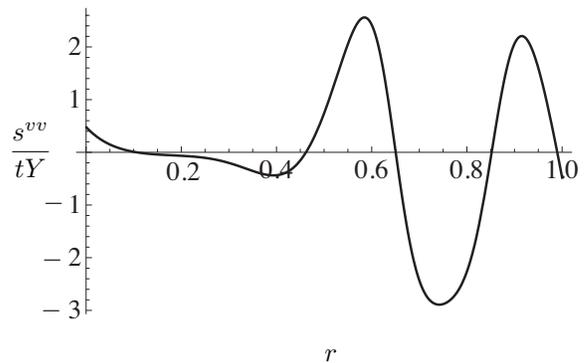}
\caption{\label{fig:stress} Nondimensionalized azimuthal stress, $s^{v v}/(tY)$, in the swelled surface of Fig. \ref{fig:tubea}, with $t=1/20$.}
\end{figure}

\section{Discussion and Conclusions}

We have presented equations that govern the design of shapes by isotropic growth of a thin elastic sheet, and solved them analytically for axisymmetric cases.  There are two relevant cases:  disk-like sheets with one boundary, and annular sheets with two boundaries.  For disk-like sheets, the boundary conditions can be satisfied locally by choosing appropriate gradients in the metric near the edges. This implies, among other things, that a negative Gaussian curvature lip must appear. For a generic annular sheet, not only must we satisfy the local boundary conditions at two boundaries, but a difficult nonlocal condition resulting from in-plane force balance.  We have found that an additional term in $\rho(u)$ linear in $u$ with adjustable coefficient can be used to satisfy this boundary condition without changing the surface dramatically. Once the assumption of axisymmetry is lifted, it is not at all clear what would be required to satisfy this last boundary condition.  

Several obstacles to swelling a shape may arise.  A chosen functional form for $\rho(u)$ may, after boundary conditions are applied, fail to satisfy $|\rho'(u)| < 1$ at one or more places on the sheet.  This implies that an axisymmetric shape of the desired form cannot satisfy the boundary conditions.  For a sufficiently thick sheet, the required prescribed metric may fail to remain positive definite at one or more places.  Thus, the shape would not be swellable, at least not by an axisymmetric target metric, at the desired thickness. This is not a surprising result, as sufficiently thick sheets will not buckle at all in response to an $\mathcal{O}(1)$ inhomogeneity in swelling.  Yet another problem-- one that we did not encounter-- may arise because our method of search for an isotropic swelling factor rests on an implicit assumption about the existence of a global conformal coordinate system.  Such a coordinate system is only guaranteed to exist locally \cite{Chern55}. Failure to find a coordinate system of this type globally suggests, again, that a particular shape may not be swellable.  Finally, we note that we have not investigated the stability of the generated surfaces, only whether axisymmetric extrema exist. The azimuthal stress $s^{v v}$ (Fig.\ \ref{fig:stress}) in the surface from Fig.\ \ref{fig:tubea} oscillates between tensile and compressive, which suggests the possibility of a wrinkling instability in the compressed regions of sufficiently thin sheets.

\section*{Acknowledgments}
We thank R. Schroll for the dynamic pressure analogy, L. Mahadevan for pointing out the role of global force and torque balance, and M. M\"{u}ller for discussions.  CDS and JAH thank the Aspen Center for Physics for its hospitality.  We acknowledge funding from the NSF through grant no.\ DMR-0846582.

\appendix

\section{Derivation of the 2D Energy}\label{3Dto2Dreduction}

Our procedure mirrors that of Efrati \emph{et al.} \cite{Efrati09JMPS}, but involves some additions.  In what follows, Latin indices run from 1 to 3, Greek indices from 1 to 2.

We begin with a three-dimensional body with elastic energy
\begin{eqnarray}\label{3Denergy}
	E&=&\frac{1}{2}\int_{\mathcal{B}} \sqrt{g} \, A^{ijkl}e_{ij}e_{kl}\\
	 &=& \frac{1}{2}\int_{\mathcal{B}} \sqrt{g} \, \left[\lambda (\mathrm{Tr}(e))^2 + 2\mu \mathrm{Tr}(e^2) \right] \, ,\nonumber
\end{eqnarray} 
using the following definitions for the elastic tensor and strain:
\begin{eqnarray}
	A^{ijkl} &\equiv& \frac{Y}{1+\nu} \left[\frac{\nu}{1-2\nu} \, g^{ij}g^{kl} + \frac{1}{2}\left(g^{ik}g^{kl}+g^{il}g^{jk}\right)\right] \, ,\nonumber \\
	2e_{ij} &\equiv& g_{ij} - \bar{g}_{ij} \, .\nonumber
\end{eqnarray}
Here $Y$ and $\nu$ are an isotropic Young's modulus and Poisson's ratio, and $\bar{g}_{ij}$ is a target metric.

If our body is a thin plate or shell, we may view it as a stack of surfaces, express quantities on each surface as expansions around the middle surface \cite{Flugge72}, and integrate along the thin dimension $z$ to obtain a two-dimensional energy.  A plate is merely a special case of a shell, in which the target metric is independent of $z$.

The position vector in the body is expressed using the midsurface immersion and unit normal
\begin{equation}
	\bR = \bX + z \bN \, ,
\end{equation}
so the metric is block-diagonal
\begin{equation}
	g_{ij} = \partial_i \bR \cdot \partial_j \bR =   \begin{pmatrix} \left(g_{\alpha\beta}\right)_{2\times2}&\bigcirc \\ \bigcirc&1\\ \end{pmatrix} \, ,
\end{equation}
and tangents to surfaces in the stack are
\begin{equation}
	\partial_\alpha \bR = \partial_\alpha \bX + z\partial_\alpha \bN = \left(\delta^\beta_\alpha - z b^\beta_\alpha\right)\partial_\beta \bX \equiv \pi^\beta_\alpha \partial_\beta \bX  \, ,
\end{equation}
where we have defined a tensor $\pi^\beta_\alpha$.  This tensor has an inverse $\rho^\alpha_\beta$ such that
\begin{equation}
	\partial^\alpha \bR \equiv \rho^\alpha_\beta \partial^\beta \bX \, , 
\end{equation}
and hence
\begin{eqnarray}
	\partial_\alpha \bR \cdot \partial_\beta \bR =& \, g_{\alpha\beta} \, &= \pi^\gamma_\alpha \pi^\delta_\beta a_{\gamma\delta} \, ,\\
	\partial^\alpha \bR \cdot \partial_\beta \bR =& \, \delta^\alpha_\beta \,\,\, &= \rho^\alpha_\gamma \pi^\gamma_\beta \, ,\\
	\partial^\alpha \bR \cdot \partial^\beta \bR =& \, g^{\alpha\beta} \, &= \rho^\alpha_\gamma \rho^\beta_\delta a^{\gamma\delta} \, .
\end{eqnarray}
With these definitions, we find that:
\begin{equation}
	\rho^\beta_\alpha = \delta^\beta_\alpha + z b^\beta_\alpha + z^2 b^\beta_\gamma b^\gamma_\alpha + \mathcal{O}(z^3) \, ,
\end{equation}
and
\begin{equation}
	\sqrt{g} = (1 - 2zH + z^2 K)\sqrt{a} \, .
\end{equation}

Moving from geometric to physical quantities, we apply the first \cite{Efrati09JMPS} Kirchhoff-Love assumption $A^{i3kl}e_{kl} =0$ 
to express our elastic energy as
\begin{eqnarray}
	A^{ijkl}e_{ij}e_{kl} &=&  \frac{Y}{1+\nu} \left[\frac{\nu}{1-\nu} \, e^\alpha_\alpha (z) e^\beta_\beta (z) +e^\alpha_\beta (z) e^\beta_\alpha (z) \right]\nonumber\\
	 &=&	A^{\alpha\beta\gamma\delta}(z) e_{\alpha\beta}(z) e_{\gamma\delta}(z) \, .
\end{eqnarray}
The second \cite{Efrati09JMPS} Kirchhoff-Love assumption $e_{i3}=0$ gives us a block-diagonal form for the target metric:
\begin{equation}
	\bar{g}_{\alpha\beta} = \begin{pmatrix} \left(\bar{a}_{\alpha\beta}\right)_{2\times2}&\bigcirc \\ \bigcirc&1\\ \end{pmatrix} \, .
\end{equation}
Finally, the expansion of the elastic tensor takes the form
\begin{equation}
\begin{split}
	A^{\alpha\beta\gamma\delta} =& \, \rho^\alpha_\kappa \rho^\beta_\lambda \rho^\gamma_\mu \rho^\delta_\nu \mathcal{A}^{\kappa\lambda\mu\nu} \\
	=& \, \mathcal{A}^{\kappa\lambda\mu\nu} \!\!\!\! \sum_{ k,l,m,n=0 }^\infty \!\!\!\! z^{k+l+m+n} (b^k)^\alpha_\kappa (b^l)^\beta_\lambda (b^m)^\gamma_\mu (b^n)^\delta_\nu \\
	=& \, \mathcal{A}^{\alpha\beta\gamma\delta} + 2z \left(\mathcal{A}^{\kappa\beta\gamma\delta}b^\alpha_\kappa + \mathcal{A}^{\alpha\beta\kappa\delta}b^\gamma_\kappa \right) + \mathcal{O}(z^2) \, ,
\end{split}
\end{equation}
where we have defined $(b^k)^\alpha_\kappa$ as the contracted product of $k$ $b$-components $b^\alpha_\beta b^\beta_\gamma \ldots b^\gamma_\kappa$ and used the definition \eqref{2Delastictensor} from the main text.

Rolling all of this together, and recalling the definition $2\varepsilon_{\alpha\beta} \equiv a_{\alpha\beta}-\bar{a}_{\alpha\beta}$, our elastic energy \eqref{3Denergy} is
\begin{eqnarray}
E &=& \frac{1}{2}\int_{\mathcal{S}} \sqrt{a} \int_{-t/2}^{t/2} dz \, \left(1 - 2zH + z^2 K\right) \\
& &\times \left[ \mathcal{A}^{\alpha\beta\gamma\delta} + 2z \left(\mathcal{A}^{\kappa\beta\gamma\delta}b^\alpha_\kappa + \mathcal{A}^{\alpha\beta\kappa\delta}b^\gamma_\kappa \right) + \mathcal{O}(z^2) \right] \nonumber\\
& &\times \left[\varepsilon_{\alpha\beta}\varepsilon_{\gamma\delta} - 2z\varepsilon_{\alpha\beta}b_{\gamma\delta} + z^2 \left(b_{\alpha\beta}b_{\gamma\delta} + \varepsilon_{\alpha\beta}b_\gamma^\kappa b_{\kappa\delta}\right) \right.\nonumber\\
& &\left. + \mathcal{O}(z^3)\right] \, ,\nonumber
\end{eqnarray}
which evaluates to \eqref{2Denergy}.

\section{Variation of the 2D Energy}\label{2Dvariation}

We consider a variation of the midsurface configuration
\begin{equation}
	\delta \bX = \delta u^\alpha \partial_\alpha \bX + \delta \zeta \bN \, ,
\end{equation}
under which the forms vary as follows \cite{DesernoNotesDG}:
\begin{eqnarray}
	\delta a_{\alpha\beta} &=& \nabla_\alpha \delta u_\beta + \nabla_\beta \delta u_\alpha - 2 b_{\alpha\beta}\delta \zeta \, , \\
	\delta b_{\alpha\beta} &=& ( \nabla_\beta \delta u^\gamma - b_\beta^\gamma \delta \zeta ) b_{\alpha\gamma} + \nabla_\alpha \left(b_{\beta\gamma}\delta u^\gamma\right) +\nabla_\alpha \nabla_\beta \delta \zeta \, .\nonumber
\end{eqnarray}  
Straightforward application of these expressions provides the useful variations
\begin{eqnarray}
	\delta \mathcal{A}^{\alpha\beta\gamma\mu} &=& -\left( \mathcal{A}^{\lambda\beta\gamma\mu}a^{\alpha\kappa} + \mathcal{A}^{\alpha\beta\gamma\lambda}a^{\mu\kappa} \right) \delta a_{\kappa\lambda} \, , \\
	\delta \sqrt{a} &=& \left(\nabla_\alpha \delta u^\alpha - 2H\delta \zeta\right)\sqrt{a} \, , \\
	\delta H &=& \frac{1}{2}\nabla_\alpha\nabla^\alpha \delta \zeta\\
	& & + \left(2H^2-K\right)\delta \zeta + \delta u^\alpha \nabla_\alpha H \, ,\nonumber
\end{eqnarray}
while a laborious calculation making use of the Peterson-Mainardi-Codazzi relations \eqref{codazzi} yields:
\begin{equation}
	\delta \left(\sqrt{a} K\right) = \nabla_\alpha \left( \sqrt{a} \left[ K \delta u^\alpha + \left(2Ha^{\alpha\beta} - b^{\alpha\beta}\right)\nabla_\beta \delta \zeta \right] \right) \, .
\end{equation}
That this last expression is a divergence, and hence only relevant on the boundaries, should be expected from the Gauss-Bonnet theorem \cite{doCarmo76}.

The equations (\ref{normalequil}-\ref{tangentialequil}) and boundary conditions (\ref{normalbc}-\ref{cornerbc}) follow from stationarity of the energy \eqref{2Denergy} with respect to arbitrary $\delta u^\alpha$ and $\delta \zeta$ in the bulk and on the boundaries, boundary-normal tangent derivatives $\partial_n \delta \zeta$ on the boundaries, and boundary-tangent tangent derivatives $\partial_l \delta \zeta$ at corners, neglecting terms of orders $t^3\| b\| ^2\|\varepsilon\|$ and $t\|\varepsilon\|^2$ after the variation.

\bibliographystyle{apsrev}

\end{document}